\documentclass [11pt, letterpaper]{article}
\pdfoutput=1 
\usepackage{fullpage}
\usepackage{amsmath}
\usepackage{amssymb}
\usepackage{graphicx}
\usepackage{braket}
\usepackage{mathrsfs}
\usepackage{wrapfig}
\usepackage{boxedminipage}
\usepackage{epsfig}
\usepackage{setspace}
\usepackage{subfigure}
\usepackage{float}
\usepackage{hyperref}
\usepackage{enumerate}
\usepackage{cite}
\usepackage{bbold}
\usepackage[font=footnotesize,labelfont=rm]{caption}

\def\Tr{{\rm Tr}}

\begin{document}

\begin{flushright}
{\tt arXiv:1807.$\_\,\_\,\_\,\_\,\_$}
\end{flushright}


\bigskip
\bigskip

\bigskip
\bigskip
\bigskip
\bigskip

\begin{center} 

{\Large\bf   Physical Generalizations of the R\'enyi Entropy}

%
%
%
%

\end{center}

\bigskip \bigskip \bigskip \bigskip

\centerline{\bf Clifford V. Johnson}

\bigskip
\bigskip
\bigskip

\centerline{\it Department of Physics and Astronomy }
\centerline{\it University of
Southern California}
\centerline{\it Los Angeles, CA 90089-0484, U.S.A.}

\bigskip

\centerline{\small {\tt johnson1}  [at] usc [dot] edu}

\bigskip
\bigskip


\begin{abstract} 
\noindent We present a new type of  generalization of the R\'enyi entropy that follows  naturally from its representation as a thermodynamic quantity. We apply it to the case of $d$--dimensional conformal field theories (CFTs) reduced on a  region bounded by a sphere. It is known how to  compute their R\'enyi entropy as  an  integral of the thermal entropy of hyperbolic black holes in $(d+1)$--dimensional anti--de Sitter spacetime.  We show how this integral fits into the framework of {\it extended} gravitational thermodynamics, and then point out the  natural generalization of the R\'enyi entropy that suggests itself in that light. In field theory terms, the new generalization employs aspects of  the physics of Renormalization Group (RG) flow to define a refined version of the reduced vacuum density matrix. For $d=2$, it can be derived directly in terms of  twist operators in field theory. The  framework presented here may have applications beyond this context, perhaps in studies of both quantum and classical information theoretic properties of a variety of systems.
\end{abstract}

\pagenumbering{gobble}

\newpage 

\pagenumbering{arabic}

\baselineskip=18pt 
\setcounter{footnote}{0}

\section{Introduction}
\label{sec:introduction}

The R\'enyi entropy~\cite{rŽnyi1961} of classical probabilistic systems is a generalization (or family of generalizations) of the Shannon entropy~\cite{shannon1971mathematical} that is of considerable interest in a wide range of fields. Given a sample of probabilities $p_i$ ($i=1\cdots N$) such that $\sum_i p_i=1$, it is defined as:
\begin{equation}
S_q=\frac{1}{1-q}\log\left(\sum_{i=1}^N p_i^q\right)\ ,
\end{equation}
where $q$ is a positive number. In the limit $q\to1$, it reduces to
$S_1=-\sum_{i=1}^N p_i\log p_i ,$
which is the Shannon entropy\footnote{This is readily demonstrated by the use of l'H\^opital's rule.}. Each value of $q$ can be considered a distinct entropic characterization of the probability sample. There is a natural analogue of all this in quantum systems as well, whether finite or infinite dimensional 
Given a density matrix $\rho=|\psi\!><\!\psi|$ in such a system (for some state $|\psi\!>$), we have R\'enyi defined as:
\begin{equation}
\label{eq:basic-renyi}
S_q=\frac{1}{1-q}\log\left[\Tr(\rho^q)\right]\ ,
\end{equation}
and in the limit $q\to1$, it becomes
$S_1=-\Tr (\rho\log \rho) ,$
the von Neumann entropy~\cite{von2018mathematical}, very familiar for measuring {\it e.g.,}  entanglement\cite{Srednicki:1993im,Bombelli:1986rw}. 

It has been noticed\cite{2011arXiv1102.2098B} that another way of thinking about the R\'enyi entropy is as a thermodynamic quantity (see also refs.~\cite{Casini:2010kt,Hung:2011nu,Klebanov:2011uf}). The density matrix  is represented as $\rho=\exp(-{\cal H}/T_0)/Z(T_0)$, where ${\cal H}$ is an appropriate Hamiltonian and $T_0$ is a temperature. The partition function $Z(T_0)={\rm Tr}\exp(-{\cal H}/T_0)$ is in the right place to make ${\rm Tr}(\rho)=1$, the analogue of $\sum_ip_i=1$ for the case of classical probabilities. We also have $Z(T_0)=\exp(-F/T_0)$ where $F(V,T)=U-TS$ is the Helmholtz free energy of the system, through which the First Law of Thermodynamics is expressed as $dF=-pdV-SdT$. Here $U$ is the internal energy and $S=-\partial F/\partial T|_V$ is  the standard (Gibbs--Boltzman) thermal entropy. Through the logarithm in the R\'enyi definition in eqn.~(\ref{eq:basic-renyi}), the system can be rewritten in terms of $F(T)$ as:
\begin{equation}
\label{eq:basic-renyi-a}
S_q=\frac{1}{\Delta T}\left[F(T_0)-F(T_0/q)\right]=\frac{1}{\Delta T}\int_{T_0/q}^{T_0} S(T)dT\ ,
\end{equation}
where $\Delta T\equiv T_0-T_0/q$. Presented this way, the case $q\to1$ obviously reduces to the thermal entropy~$S$. In summary, the R\'enyi entropy is an appropriately normalized measure of the free energy difference between the system at temperature $T_0$ and the system at temperature $T_0/q$, assuming the volume~$V$ is held fixed. 

In this paper, we point out  and explore  a generalization of R\'enyi that is (to our knowledge) quite different in origin and character from generalizations in the literature, and that could well have  some useful applications. The whole discussion so far had $V$ and its conjugate variable $p$ essentially turned off. Physics suggests that it is quite natural to turn them on, depending upon the context\footnote{Particle number and conjugate chemical potential could also be included too.}. Then instead of just ``quenching'' the temperature, one can also see how the free energy changes by adjusting $V$ or~$p$ in a similar way as was done for $T$. In this way, we will define an extension of R\'enyi denoted $S_{q,b}$, where $b$ is analogous to (but logically independent from) the parameter $q$. It will be defined with an appropriate normalization such that when $q,b\to1$ we recover the von Neumman entropy. 

It seems more natural to focus on the pressure $p$, as it is the intensive variable analogous to~$T$ (although it is an easy exercise to present things in terms of $V$). Focussing on $p$, we'll see that the natural free energy is not  the Helmholtz free energy $F$, but the Gibbs free energy $G(p,T)=F+pV$. Evidently, the partition function in play here is the grand partition function $Z(T,p)={\rm Tr}\exp(-({\cal H}+pV)/T_0)$.

Our way of extending R\'enyi from $S_q$ to our new quantity $S_{q,b}$ may (with the right interpretation) be of interest not just in the quantum context just outlined, but also in classical information theory. It will (for generic choices of $b$) possibly  fall outside the usual definitions of what is considered desirable\cite{rŽnyi1961,shannon1971mathematical} for a   measure of entropy, but we will leave that issue for future exploration. The thermodynamic motivation presented above is highly suggestive that  $S_{q,b}$'s properties should be further explored. Moreover, as we will explain next, in the context of $d$--dimensional conformal field theories, the extension will turn out to be extremely natural. This is because there is a dual gravitational system for which the {\it extended} gravitational thermodynamics (in the sense of ref.~\cite{Kastor:2009wy}, where $p$ and~$V$ are variables) provides just the right structures to accommodate our extension. In ref.~\cite{Casini:2011kv} it was shown how to relate the entanglement entropy of the CFTs reduced on a round ball of radius $R$ to  the thermal entropy of a hyperbolically sliced AdS$_{d+1}$. This spacetime is really just a special (massless) case of a black hole.  The work of ref.~\cite{Hung:2011nu}  built on that  computation by showing how to compute the R\'enyi entropy,  by integrating the entropy $S(T)$ of  the neighbouring $M\neq0$ black hole solutions. Those works were at fixed cosmological constant.  Recently, ref.~\cite{Johnson:2018amj} showed that the physics of the hyperbolic black holes in AdS$_{d+1}$ {\it with dynamical cosmological constant}~$\Lambda$ has direct interpretations in the CFT$_{d}$.
%
%
In this extended thermodynamics, dynamical $\Lambda$ supplies a pressure~$p$ for the black hole dynamics, and processes that change $p$ correspond to perturbations of the CFT that include excitations and  generalized flows, changing the number of degrees of freedom\footnote{A subset of those flows are Renormalization Group flows, and, amusingly, various generalized $c$--theorems emerge as consequences of the Second Law of Thermodynamics.}. As we will see, the generalization of the R\'enyi entropy, $S_{q,b}$, that we define for these systems will  utilise such  departures into the infrared (IR) and ultraviolet (UV) in a natural way. It will be very analogous to how standard R\'enyi entropy is computed using departures away from the central temperature of interest, $T_0$. For $d=2$ we will be able to be quite explicit about how it extends the usual computations in conformal field theory.

\section{Entanglement Entropy of CFT}
\label{sec:entanglement-entropy}
Our model  system will be a $d$--dimensional conformal field theory in Minkowski space, and there is a subregion (called $B$) which  is a round ball of radius $R$, whose surface is a round sphere~$\mathbb{S}^{d-2}$. The reduced density matrix of interest is $\rho_{\rm v}={\rm Tr}_B (\rho)$ where $\rho=|0\!\!><\!\!0|$.  In ref.~\cite{Casini:2011kv} it was shown that in such a case, a conformal transformation can be used to map the system to a hyperbolic spacetime $\mathbb{R}\times\mathbb{H}^{d-1}$. The entanglement entropy maps to  a {\it thermal} entropy ({\it i.e.,} that of Boltzmann and Gibbs). This is  because a conformal map can be constructed that takes the reduced density matrix $\rho_{\rm v}$ and relates  it to a thermal one {\it via}:
\begin{equation}
\label{eq:conformal-to-thermal}
\rho_{\rm v}=U^\dagger \left( \frac{e^{-\beta {\cal H}}}{Z} \right)U\ , 
\end{equation}
where $\beta^{-1}=T_0=1/2\pi R$, and $U$ is the representation of the conformal map acting in state space. The Hamiltonian ${\cal H}$ generates  time translations in the  hyperbolic spacetime. The characteristic curvature length scale of the $\mathbb{H}$ factor is $L_0$, which is identified with~$R$. This spacetime can be written as the boundary  of a $(d+1)$--dimensional AdS spacetime with hyperbolic slicing:
\begin{equation}
\label{eq:ads-hyperbolic}
ds^2=-\left(\frac{\rho^2}{L_0^2}-1\right)d\tau'^2+
  \frac{d\rho^2}{\left(\frac{\rho^2}{L_0^2}-1\right)}+\rho^2\left(du^2+\sinh^2(u)d\Omega_{d-2}^2\right)\ ,
\end{equation}
with a radial coordinate $\rho$ such that  $L_0\leq\rho\leq+\infty)$. The coordinate $u$ has the range $0\leq u\leq +\infty$. There is a hyperbolic  horizon at $\rho=\rho_+=L_0$, which results in  this spacetime having a temperature $T_0=1/2\pi L_0$.  As we shall see, this is a special massless black hole that is part of a larger family of black holes to be discussed. The thermal entropy is given by the Bekenstein--Hawking formula\cite{Bekenstein:1973ur,Hawking:1974sw,Hawking:1976de}  relating it to the area of the horizon divided by $4G_{\rm N}$ (where $G_{\rm N}$ is Newton's constant in $(d+1)$ dimensions)\cite{Emparan:1999pm,Emparan:1999gf}:
\begin{equation}\label{eq:temp-ent}
  S^{(M=0)}=\frac{w_{d-1}L_0^{d-1}}{4G_{\rm N}}=\frac{L_0^{d-1}}{4G_{\rm N}}
 \Omega_{d-2}
  \int_{0}^{+\infty} du\,
  \sinh^{d-2}(u)\ ,
\end{equation}
where $w_{d-1}$ is the volume ({\it i.e.,} surface area) of the hyperbolic plane with radius one and in the final term we have written it more explicitly, with $\Omega_{d-2}$ the volume ({\it i.e.,} surface area) of a  unit sphere $\mathbb{S}^{d-2}$. 

In mapping to an entanglement entropy, the divergent $u$ integral must be cut off at some~$u_{\rm max}$. In terms of the variable $x=\sinh(u)$, this cutoff is $x_{\rm max}=\sqrt{(R/\epsilon)^2-1}$, and  the regulated entropy is given by\cite{Casini:2011kv}:
\begin{equation}
\label{eq:entanglement}
S_{\rm EE}=
  \left(\frac{2\Gamma(d/2)\Omega_{d-2}}{\pi^{d/2-1}}\right)
  a^*_d  \int_{0}^{x_{\rm max}} dx\,  \frac{x^{d-2}}{\sqrt{1+x^2}}\ ,
 \end{equation}
where the $L_0$ dependent factor was written in terms of $a^*_d$, the generalized central charge\cite{Casini:2011kv}:
\begin{equation}\label{eq:a-function}
a^*_d=\frac{\pi^{d/2-1}}{8 \Gamma(d/2)}\frac{L_0^{d-1}}{G_{\rm N}}\ ,
\end{equation}
where $L_0$ is the AdS radius, $G_{\rm N}=(4\pi)^{-1}L_p^{d-1}$ is Newton's constant, and $L_p$ is the Planck length, in $d+1$ dimensions. For each $d$, the ratio $L_0/L_p$ can be written in terms of purely field theory quantities using the holographic duality. For example, $12\pi L_0/L_p=c$, the standard central charge in $d=2$,  and $S_{\rm EE}=(c/3)\log (\ell/\epsilon)$, where $\ell=2R$, the length of the interval that is the width of the ``ball'' in 1 spatial dimension. (We will discuss more about $d=2$ in subsection~\ref{sec:d=2}.) For higher~$d$, the quantities $a_d^*$ are a generalization\cite{Cardy:1988cwa,Casini:2011kv} of $c$. 

In ref.~\cite{Johnson:2018amj}, the use of this map was extended to the case of dynamical cosmological constant~$\Lambda$, interpreted as a pressure $p=-\Lambda/8\pi G_{\rm N}$. Dynamical $p$  (and its conjugate volume $V$) will allow for the physics to explore flows of the CFT into the IR and UV. This will  be discussed  in section~\ref{sec:extended-thermo}.

\section{R\'enyi Entropy of CFT}
As already mentioned, the metric in eqn.~(\ref{eq:ads-hyperbolic}) is a special case of a larger family of metrics representing hyperbolic black holes\cite{Birmingham:1998nr}\footnote{Four dimensional versions of these hyperbolic black holes first appeared in the literature as ``topological black holes'' in refs.\cite{Mann:1996gj,Vanzo:1997gw,Brill:1997mf,Emparan:1998he}, accompanied by some discussion of their thermodynamics.}:
\begin{equation}
ds^2=-{\overline V}(\rho)d\tau'^2+\frac{d\rho^2}{{\overline V}(\rho)}+\rho^2\left(du^2+\sinh^2(u)d\Omega^2_{d-2}\right)\ ,
\end{equation}
where
\begin{equation}
{\overline V}(\rho)=\frac{\rho^2}{L^2}-\frac{\mu}{\rho^{d-2}}-1\ .
\end{equation}
We are using a general $L$ here as the AdS length scale for allow for the possibility of scales different from~$L=L_0$ in later sections. 
The horizon radius $\rho=\rho_+$ is defined by the vanishing  ${\overline V}(\rho_+)=0$, while the factor $\mu$ is proportional to the mass of the black hole:
\begin{equation}
\label{eq:the-mass}
M(\rho_+,L)=\left(\frac{(d-1)w_{d-1}}{16\pi G_{\rm N}}\right)\mu=\left(\frac{(d-1)w_{d-1}}{16\pi G_{\rm N}}\right)
  \rho_+^{d-2}\left[\left(\frac{\rho_+}{L}\right)^2-1\right]\,,
\end{equation}
where $w_{d-1}$ is the volume of the hyperbolic space with radius one (whose regularization was discussed in section~\ref{sec:entanglement-entropy}). 
The entropy and temperature are given by
\begin{eqnarray}
\label{eq:the-entropy-temperature}
S=\frac{w_{d-1}\rho_+^{d-1}}{4G_{\rm N}}\quad{\rm and}\quad
T=
  \frac{(d-2)}{4\pi}\frac{1}{\rho_+}
  \left[\frac{d}{(d-2)}\left(\frac{\rho_+^2}{L^2}\right)-1\right]\ .
\end{eqnarray}
The key observation at this point is that one can quite readily compute the $q$th power of the reduced density matrix {\it via} the conformal map ({\it c.f.} eqn.~(\ref{eq:conformal-to-thermal})):
\begin{equation}
(\rho_{\rm v})^q=U^\dagger \left( \frac{e^{-\beta q {\cal H}}}{Z^q} \right)U\ , 
\end{equation}
and since ${\rm Tr}[ \exp(-\beta {\cal H})]=Z(T_0)$, the R\'enyi entropy can be written as shown earlier in eqn.~(\ref{eq:basic-renyi-a}), repeated here for convenience:
\begin{equation}
S_q=\frac{1}{1-q}\log\Tr(\rho_{\rm v}^q)=-\frac{F(T_0)-F(T_0/q)}{T_0-T_0/q}=\frac{1}{T_0-T_0/q}\int_{T_0/q}^{T_0}S(T) dT\ , 
\end{equation}
 \begin{wrapfigure}{l}{0.45\textwidth}
\centering
\includegraphics[width=0.4\textwidth]{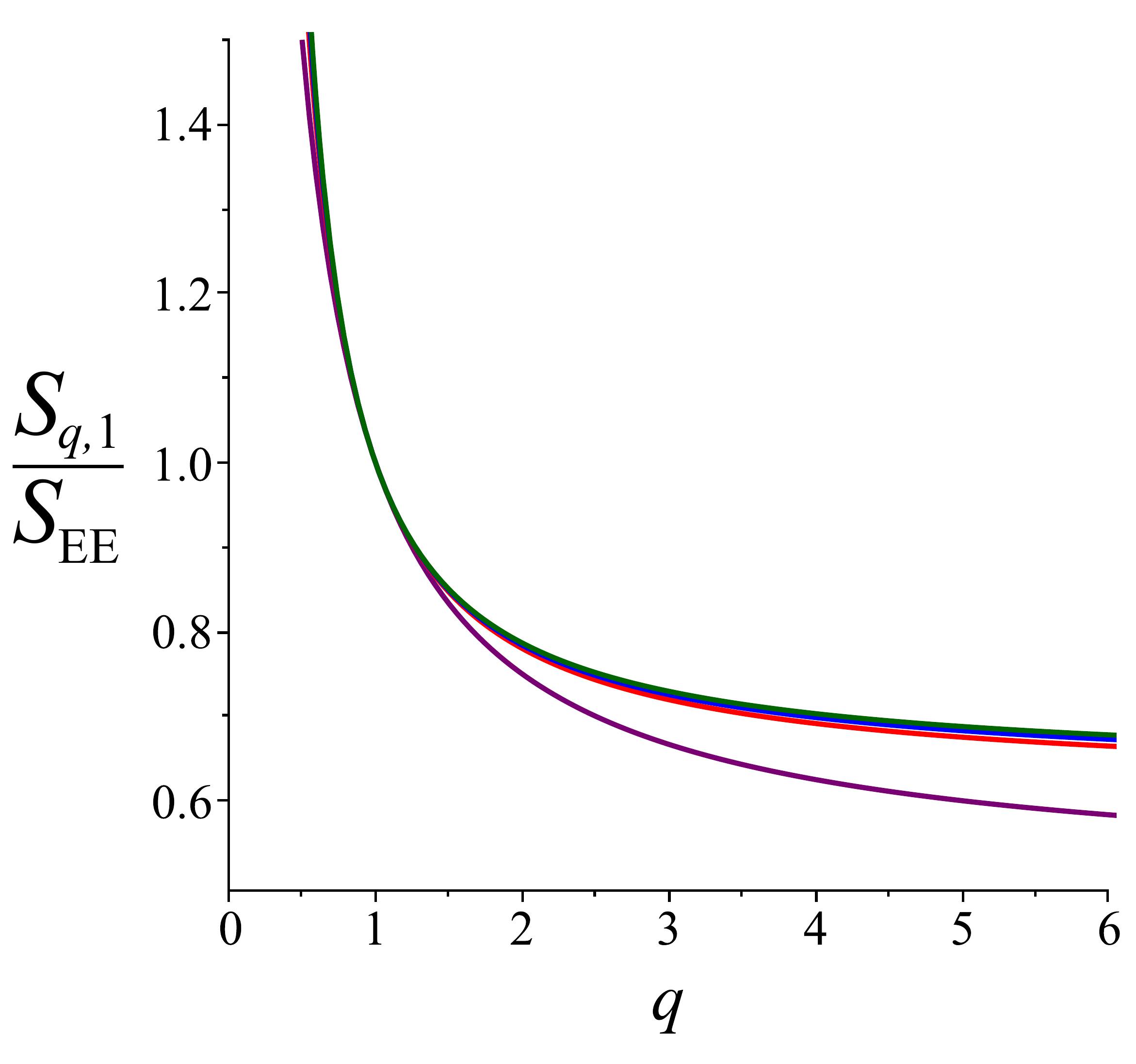}
\caption{\label{fig:renyi-vs-q} The ratio of the R\'enyi entropy to the entanglement entropy, $S_q/S_{\rm EE}$, as a function of $q$, for $d=2$ (lowest), 3, 4 and 10. All the curves pass through unity at $q=1$.}
\end{wrapfigure}
where the Helmholz free energy $F(T)\equiv - T\log Z(T)$ and  $S(T)=-\partial F(T)/\partial T$. (Again, note that in this form 
it is quite straightforward to see that the limit $q\to1$ gives $S_1=S$.) Finishing the computation,  the R\'enyi entropy can be written as:
\begin{equation}
S_q= \frac{q}{q-1}\frac{1}{T_0}\int_{\bar{x}}^1 S(x)\frac{dT}{dx}dx\ ,
\end{equation}
where $x=\rho_+/L$, which is equal to 1 for the $M=0$ black hole, where $T=T_0=1/2\pi L_0$, and  $x=\bar{x}$ at $T=T_0/q=1/2\pi L_0 q$. A little algebra using eq.~(\ref{eq:the-entropy-temperature}) shows that
$\bar{x}=(1+\sqrt{1+q^2 d(d-2)})/qd$. From here, since the entropy and temperature are known  functions of  $x$ given in eq.~(\ref{eq:the-entropy-temperature}) the following is the result for the {\it regulated} R\'enyi entropy of the conformal field theory:
\begin{equation}
S_q= \frac{q}{q-1} S_{\rm EE}\left(2- \bar{x}^{d-2}(1+\bar{x}^2)\right)\ ,
\end{equation}
where $S_{\rm EE}$ is the $\epsilon$--cutoff entanglement ({\it i.e.,} von Neumann) entropy  defined in eqn.~(\ref{eq:entanglement}). In the next section $S_q$ will be renamed $S_{q,1}$, in the light of the existence of a larger structure. In figure~\ref{fig:renyi-vs-q}, some examples of the behaviour of $S_q/S_{\rm EE}=S_{q,1}/S_{\rm EE}$ as a function of $q$ are shown.

\section{Beyond  R\'enyi Entropy}
\subsection{Extended Thermodynamics}
\label{sec:extended-thermo}
The black holes discussed so far arise from enlarging the framework  from just hyperbolic AdS (the $M=0$ case), while keeping fixed the AdS scale at $L=L_0$. This allowed for a natural setting in which to compute the  R\'enyi entropy of the CFT. In fact, the framework can be  further enriched in a quite natural way. The key is to go from the ordinary black hole thermodynamics to the ``extended" framework where the cosmological constant $\Lambda=-d(d-1)/(2L^2)$ is dynamical. This defines a pressure $p=-\Lambda/8\pi G$ and a conjugate volume $V$.  The result is a family of hyperbolic black holes of mass $M$, temperature~$T$ and entropy $S$ (given in the previous section), at pressure $p$ given by replacing $L$ in all those equations  according to:
\begin{equation}
p=\frac{d(d-1)}{16\pi G}\frac{1}{L^2}\ .
\end{equation}
The mass of a black hole is identified with the enthalpy\cite{Kastor:2009wy}: $M=H(S,p)=U+pV$, and the First Law can be written as $dH=TdS+Vdp$.
The volume is given by:
\begin{equation}
\label{eq:volume}
V=\left.\frac{\partial H}{\partial p}\right|_S=\frac{w_{d-1}}{d}\rho_+^d\ .
\end{equation}
In ref.~\cite{Johnson:2018amj} the physical meaning of this larger framework was uncovered. Fixed $p$ processes keep one within the conformal field theory, but deform the state. Processes which change $p$ are readily interpreted as changing the number of degrees of freedom of the CFT as measured by the entanglement entropy, since the $a^*_d$  depends upon $L$. Focussing on the $(p,V)$ plane, for example, the $M=0$ sector corresponds to the special line $p=\kappa/V^{2/d}$, where $\kappa$ is a constant. Every point on that line can be mapped to the ball--reduced conformal field theory. Associating a given point with radius $R$, points at higher values of $p$ are deeper into the infrared (IR) while points at lower $p$ are toward the ultraviolet (UV). In fact, the simple irreversible process by which the system cools by heat outflow to a colder reservoir, reducing its volume and moving up the $M=0$ curve  to higher~$p$  was argued\cite{Johnson:2018amj} to be {\it exactly} renormalization group (RG) flow in the CFT. Theorems about the reduction of the degrees of freedom along RG flow are, in this framework, simply consequences of the Second Law of Thermodynamics.

\begin{figure}[h]
\centering
\subfigure[]{\centering\includegraphics[scale=0.65]{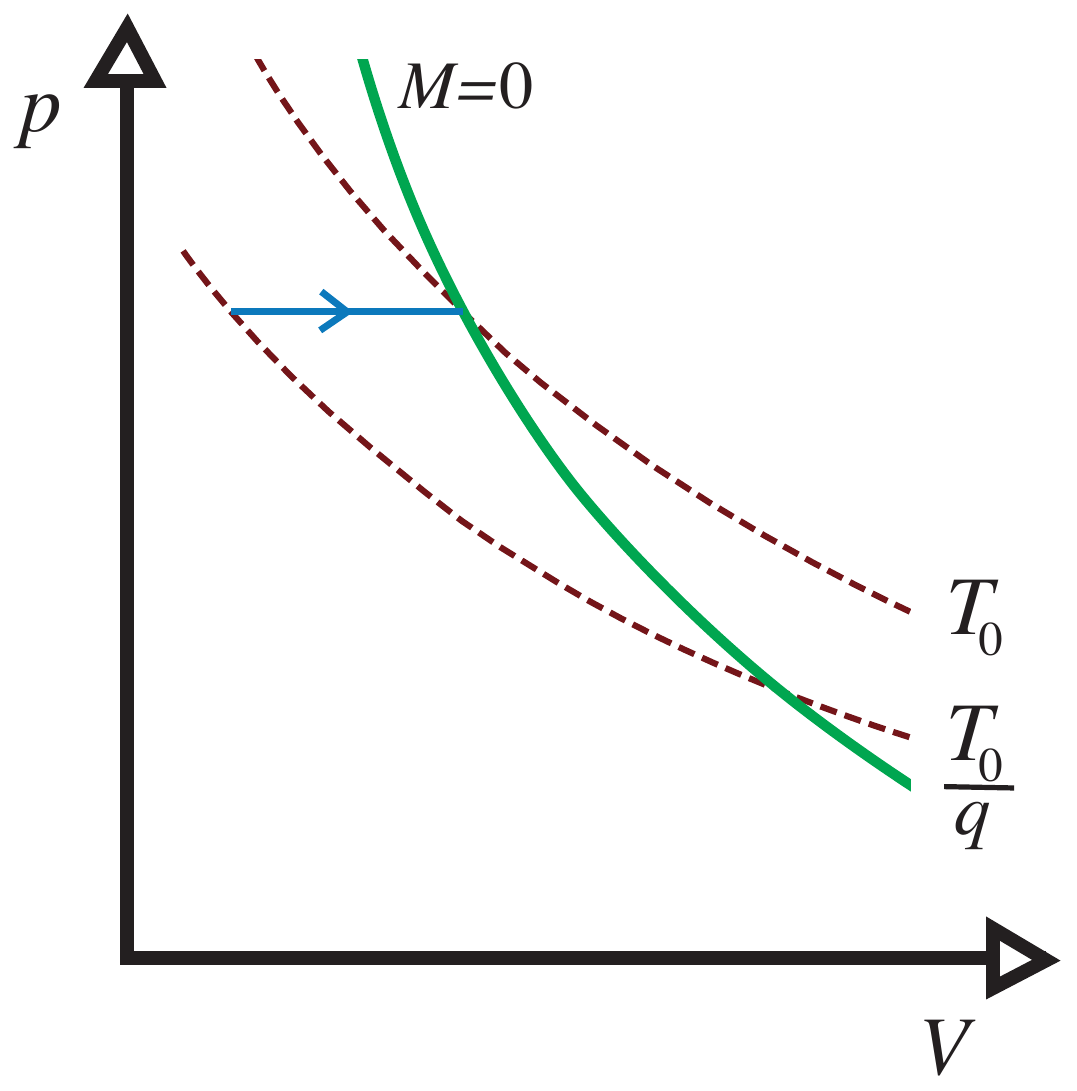}
\label{fig:pv-plane-a}}
\hspace{50pt}
\subfigure[]{\centering\includegraphics[scale=0.65]{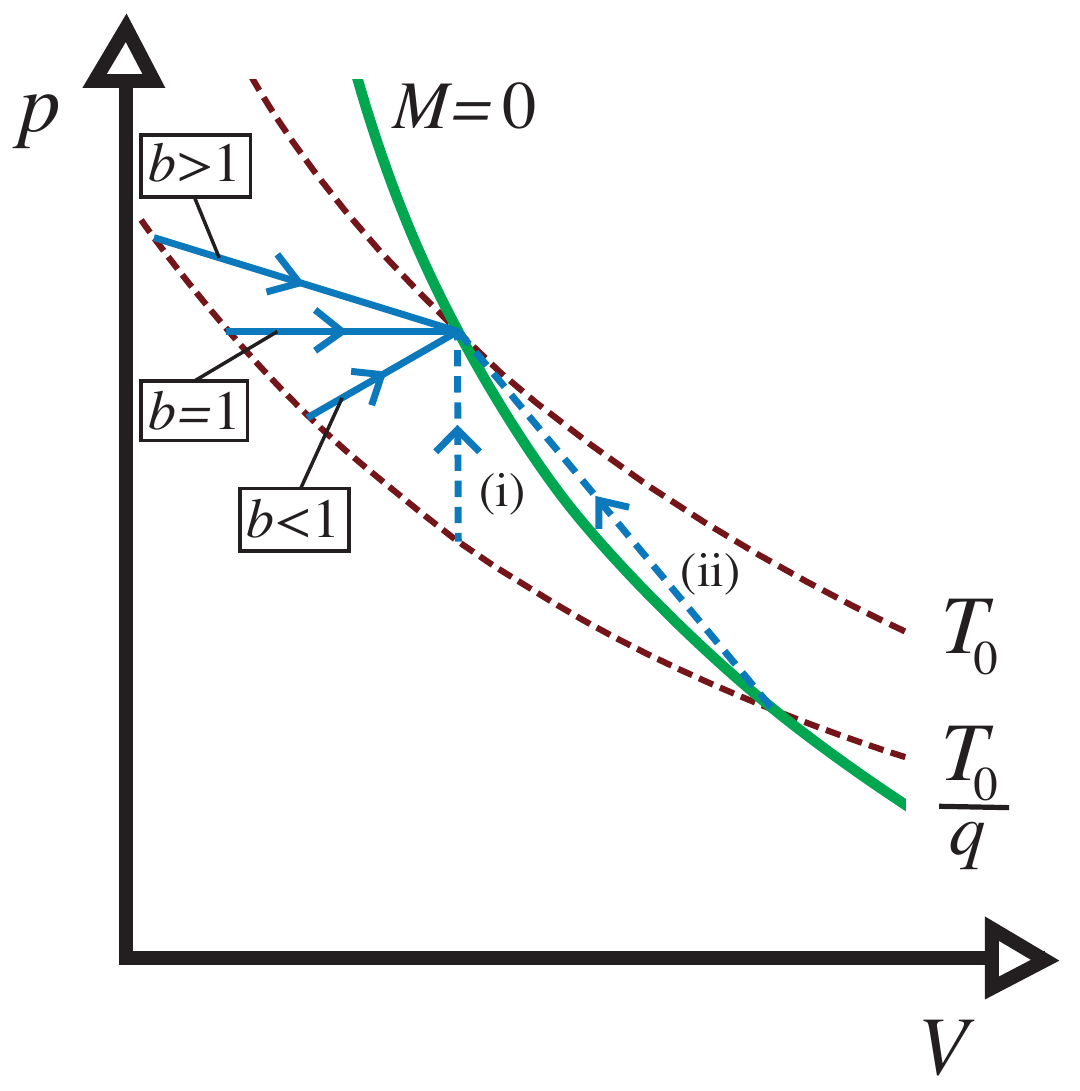}
\label{fig:pv-plane-b}
}
\caption{Paths in the $(p,V)$ plane discussed in the text for some cases with $q>1$. The solid (green) curve is  the $M=0$ curve of conformal field theories. The two dotted (red) curves are isotherms, one at $T_0=1/2\pi L_0$ and the other at $T_0/q<T_0$.  Panel~(a) shows the straight horizontal integration path between the  isotherms that defines the basic R\'enyi entropy. Since the object being integrated, the Gibbs free energy, is a thermodynamic potential, the integration only depends on the endpoints of the path. Panel~(b) additionally shows other straight integration paths that define generalizations of the R\'enyi entropy through a parameter $b$.  Paths (i) and (ii) are two special cases discussed in the text. Note that analogous diagrams can be drawn for the $q<1$ case.}\label{fig:pv-plane}
\end{figure}

Let us understand the R\'enyi entropy computation of the last section in this enlarged framework. The first observation is that the Helmholz free energy $F=U-TS$  of the ordinary thermodynamics maps to the Gibbs free energy $G=U+pV-TS=H-TS$ in the extended thermodynamics\cite{Johnson:2014yja}. It is going to be useful for us later on, and so let us compute it from  eqs.~(\ref{eq:the-mass},\ref{eq:the-entropy-temperature}):
\begin{equation}
\label{eq:gibbs}
G=-\frac{w_{d-1}}{16\pi G_N} \rho_+^{d-2}\left(1+\frac{\rho^2_+}{L^2}\right)\ .
\end{equation}
Although the natural variables for Gibbs are $p$ and $T$, we'll leave the dependence on them implicit here.
Now in terms of Gibbs, the First Law is:
\begin{equation}
\label{eq:first-law-gibbs}
dG=-Vdp-SdT\ .
\end{equation}
Since everything was done at constant pressure $p_0$ (set by $L_0$) we had for the R\'enyi entropy:
\begin{equation}
\label{eq:renyi-as-integral}
S_q=-\frac{[G(p_0,T_0)-G(p_0,T_0/q)]}{T_0-T_0/q} =  \frac{1}{(T_0-T_0/q)}\int_{T_0/q}^{T_0}S(T)dT\ .
\end{equation}
From figure~\ref{fig:pv-plane-a} this can be seen as  an integral along the horizontal (blue) line (at $p=p_0$) between isotherms at $T_0/q$ and $T_0$.

 In fact, since $G$ is a thermodynamic potential,  we could have carried out that integral {\it along any path in the $(p,V)$ plane}. It is actually more useful to simply interpret  the R\'enyi entropy as the magnitude of the (Gibbs) free energy difference between the system at temperature $T_0$ and the system at temperature $T_0/q$, divided by the temperature difference.

\subsection{Extending R\'enyi}
\label{sec:beyond}
Thinking of R\'enyi in the way just  mentioned makes it extremely natural to take the next step. There's really no reason to restrict to  temperature differences at  fixed $p=p_0$ (horizontal moves), especially in light of the fact that the entire $(p,V)$ plane has interpretation in the conformal field theory. R\'enyi would appear to be part of a much larger family of quantities $S_{q,b}$  defined  by changing $p$ as well, where  we move from~$p_0$ to~$b^2p_0$ by changing $L_0$ to $L_0/b$ where $b$ is a positive real number, just like~$q$.  Our new ``entropy" would again be the magnitude of the Gibbs free energy difference between points, appropriately normalized. Example points (and associated, optional, straight paths) are shown in figure~\ref{fig:pv-plane-b}. (Some of them are interesting special cases we will examine in subsection~\ref{sec:special-cases}.)

The question arises as to what the appropriate normalization must be. The fact is, it follows naturally from the First Law for $G$ given in eqn.~(\ref{eq:first-law-gibbs}). Integrating to get the difference gives:
\begin{equation}
\label{eq:big-entropy}
S_{q,b}= -\frac{G(p_0,T_0)-G(b^2p_0,T_0/q)}{(\Delta T+V_0\Delta p/S_0)} =  \frac{1}{(\Delta T+V_0\Delta p/S_0)}\int_{T_0/q}^{T_0}S(T)\left(1+\frac{V}{S}\frac{dp}{dT}\right) dT\ ,
\end{equation}
where 
\begin{equation}
\label{eq:deltas}
\Delta T\equiv T_0-T_0/q \ ,\quad  \quad \Delta p \equiv \frac{d(d-1)}{16\pi G_N}\frac{1}{L_0^2}\left(b^2-1\right) \ ,\quad{\rm and}\quad \frac{V_0}{S_0}= \frac{4G_N L_0 }{d}\ .
\end{equation}
Notice that this means that when the parameters $q,b\to 1$ we get $S_{1,1}=S$, so our generalized ``entropy'' will again reduce to the entanglement entropy $S_{\rm EE}$ after using the map of section~\ref{sec:entanglement-entropy}. Rather nicely, we see that there is even a family of non--trivial quantities when  $q=1$, defined for $\Delta T=0, \Delta p\neq0$ since the Gibbs free energy difference is still non--zero, and can be written  as an integral by instead using~$p$ as the integration variable  on the right hand side of equation~(\ref{eq:big-entropy}):
\begin{equation}
\label{eq:big-entropy-a}
S_{1,b}= -\frac{G(p_0,T_0)-G(b^2p_0,T_0)}{V_0\Delta p/S_0} =  \frac{S_0}{V_0\Delta p}\int_{p_0}^{b^2p_0}V(p) dp\ .
\end{equation}
Because of the factor $S_0/V_0$, these $S_{1,b}$ again reduce to $S_{\rm EE}$ as $b\to1$. It is interesting to note that the quantity $V_0/S_0\times S_{1,b}$, written as a $p$--integral of $V(p)$ is a direct analogue of writing R\'enyi ($S_{q,1}$) as a $T$--integral of $S(T)$, as in eqn.~(\ref{eq:renyi-as-integral}). This might lend some support to the idea\cite{Johnson:2014yja} that, in general, the thermodynamic volume is also a natural geometric measure of an aspect of the theory, although it remains to be seen exactly what that aspect is\footnote{There are some suggestions that it could be a measure of quantum complexity, see ref.~\cite{Couch:2016exn}. It is also connected to the geometry discussed in connection to  conjectures about ``subregion complexity"~\cite{Alishahiha:2015rta}.}.

Putting everything together we have:
\begin{equation}
\label{eq:gibbs-difference}
S_{q,b}= -\frac{q}{T_0}\frac{[G(p_0,T_0)-G(b^2p_0,T_0/q)]}{(q-1+q(d-1)(b^2-1)/2)} \ ,
\end{equation}
which, after regularizing as before, gives:
\begin{equation}
\label{eq:general-entropy}
S_{q,b}=S_{\rm EE}\cdot  \frac{q}{2}\frac{[2-\left(\frac{\bar{x}}{b}\right)^{d-2}(1+\bar{x}^2)]}{[q-1+q(d-1)(b^2-1)/2]} \ ,
\quad {\rm with} \quad
\bar{x}=\frac{1}{qbd}\left(1+\sqrt{1+d(d-2)q^2b^2}\right)\ .
\end{equation}
For later use, we note that $\bar{x}\to\sqrt{(d-2)/d}$,  for $q$ or $b \to\infty$, and 
$\bar{x}\to 2/qbd$ , for $q$ or $b \to 0$ ,
and so 
\begin{equation}
\label{eq:large-q}
\lim_{q\to\infty} \left(\frac{S_{q,b}}{S_{\rm EE}}\right) = \frac{2}{2+(d-1)(b^2-1)}\left(1-\frac{1}{b^{d-2}}\frac{d-1}{d}\left(\frac{d-2}{d}\right)^{\frac{d-2}{d}}\right)\ ,
\end{equation}
and 
\begin{equation}
\label{eq:small-q}
\lim_{q\to0} \left(\frac{S_{q,b}}{S_{\rm EE}}\right) = \frac{1}{2}\left(\frac{2}{d}\right)^d\left(\frac{1}{b}\right)^{2d-2}\frac{1}{q^{d-1}}\ .
\end{equation}
A few remarks are in order:

 {\bf (1)} The first is about our choice of parameter, $b$, for the generalization. Perhaps a choice more closely analogous to $q$ would have been $r=1/b^2$, so that the new pressure was $p_0/r$. On the other hand, the lesson learned from ref.~\cite{Johnson:2018amj} is that $b$, as a ratio of AdS length scales, is a good guide to some of the physics of the CFT. For example, when comparing the entanglement of two CFTs along the $M=0$ curve in figure~\ref{fig:pv-plane}, it gives the relative sizes of the cutoffs.

\begin{figure}[h]
\centering
\subfigure[]{\centering\includegraphics[scale=0.3]{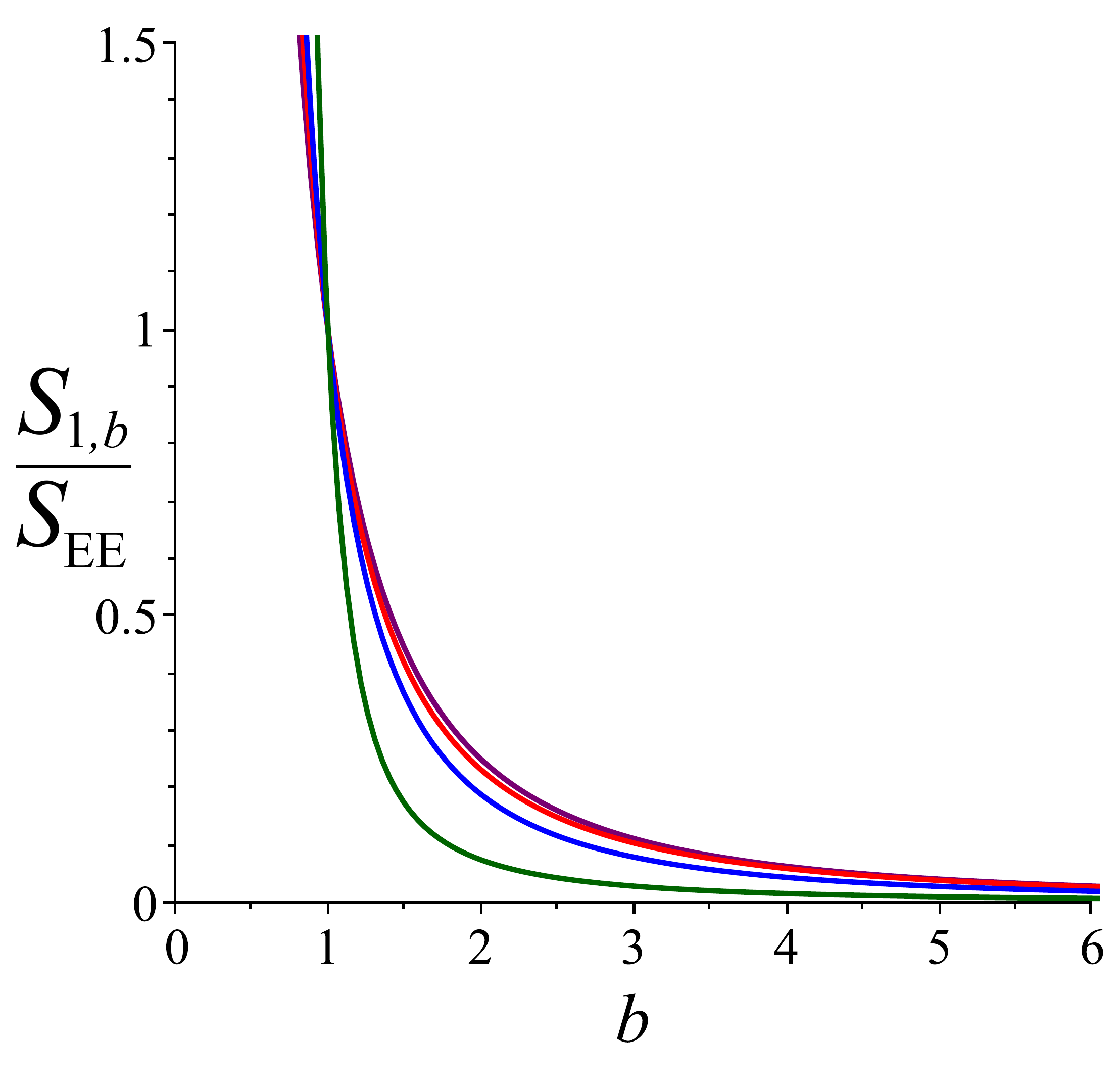}
\label{fig:renyi-fixed-q}}
\hspace{30pt}
\subfigure[]{\centering\includegraphics[scale=0.3]{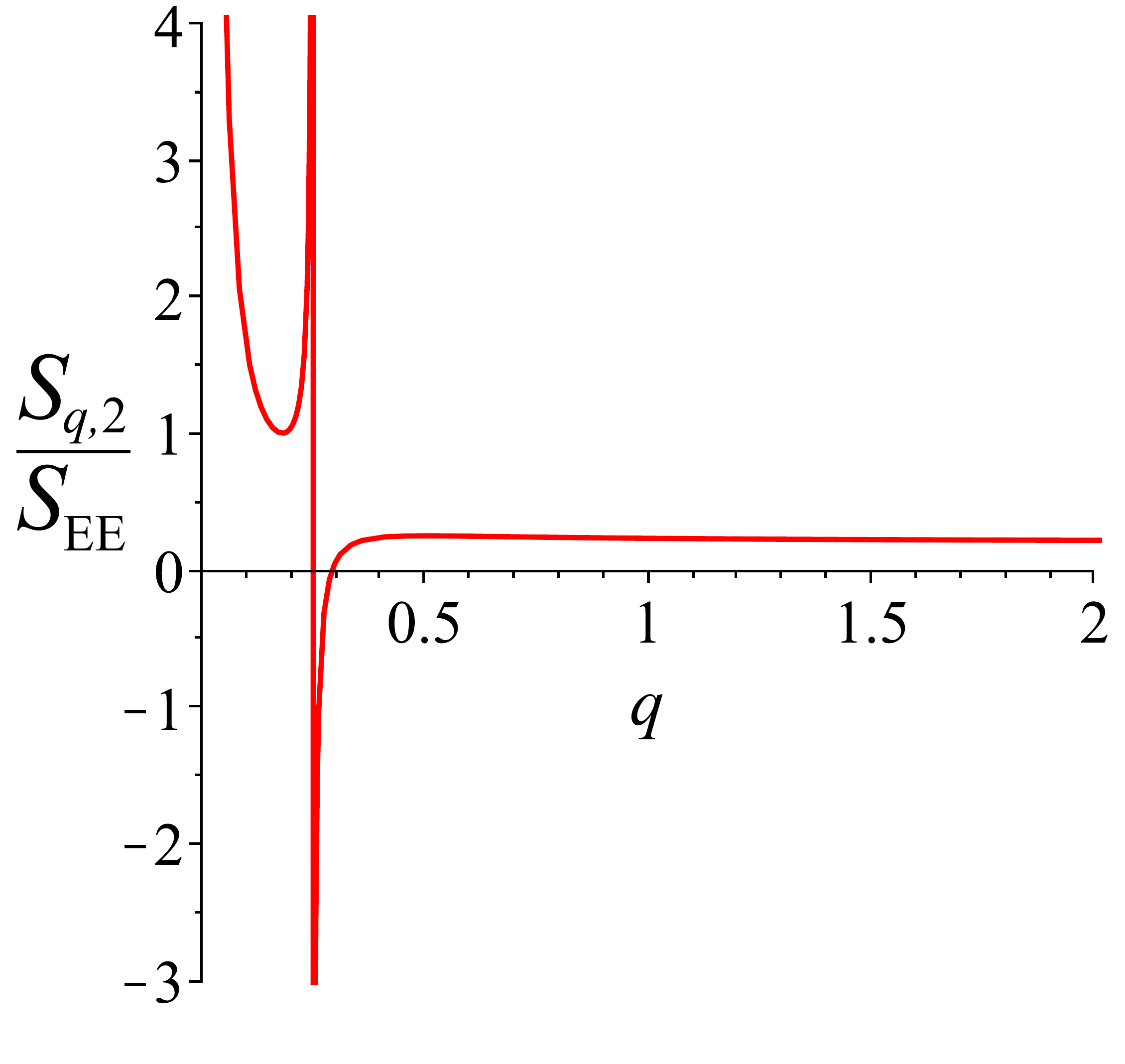}
\label{fig:renyi-fixed-b}
}
\caption{The ratio $S_{q,b}/S_{\rm EE}$ for $d=2$ (highest), 3, 4, and 10 (lowest). (a) The case of $q=1$, while varying $b$, supplying a striking analogue of the R\'enyi case $S_{q,1}$; (b) The singular case of varying  $q$ with  fixed $b\neq1$ (here $b=2$, $d=3$),  discussed more in the text.}\label{fig:renyi-plots-a}
\end{figure}

{\bf (2)} The second remark is that so far $q$ and $b$ are completely independent in general. Moreover they are on equal footing. So while we can think of $b$ as a one--parameter deformation of the R\'enyi entropy, fixing it to particular values and studying the $q$--dependence, we can also fix $q$ and look at the $b$--dependence. In particular, $q=1$ is of interest since we then have a $b$ dependent deformation of the entanglement entropy that is entirely analgous to (but different from) the R\'enyi entropy.  For $d=2$ we have 
\begin{equation}
S_{1,b}=\frac{c}{3}\left(\frac{1}{b^2}\right)\log\left(\frac{\ell}{\epsilon}\right)\ ,\quad {\it vs.}\quad S_{q,1}=\frac{c}{6}\left(1+\frac{1}{q}\right)\log\left(\frac{\ell}{\epsilon}\right) \ ,
\end{equation}
 with more complicated $b$ dependence in $S_{1,b}$ for higher $d$. 
 %
 %
 Figure~\ref{fig:renyi-fixed-q} shows the ratio $S_{1,b}/S_{\rm EE}$ as a function of $b$, for $d=2,3,4$ and 10.  It is worth comparing it to the samples of the R\'enyi entropy, $S_{q,1}/S_{\rm EE}$, displayed in figure~\ref{fig:renyi-vs-q}. For each dimension the functions all fall off faster from unity for the new quantity, and the $d$ dependence of this rate of fall off is precisely reversed. A key difference is that for all $d$, $S_{1,b}/S_{\rm EE}$ vanishes for $b\to\infty$  as  $2/(d-1)b^2$, 
as opposed to the finite constant in eq.~(\ref{eq:large-q}) (with $b=1$) that the $S_{q,1}/S_{\rm EE}$ asymptote to. The $b\to0$ divergence is $S_{1,b}\sim (b^2)^{1-d}$ instead of $S_{q,1}\sim q^{1-d}$ (see eqn.~(\ref{eq:small-q})).

The general $q$--asymptotic behaviour of $S_{q,b}$ can be read off from eqns.~(\ref{eq:small-q}) and~(\ref{eq:large-q}). Recall that the small $q$ behaviour of the standard R\'enyi entropy tells us about $\log {\rm Tr}(\mathbb{1})=\log D$, where~$D$ is a measure of the (infinite) size of the density matrix $\rho_{\rm v}$.  Now we see that in the presence of~$b$, the $\log D$ figure is {\it reduced} by a factor $b^{2d-2}$, perhaps indicating  that the effective number of degrees of freedom has been reduced\footnote{This is consistent with the fact that $b>1$ is a measure of how far into the IR we have probed. (For $b<1$ we see the converse, an enhancement).}.   In the  $q\to\infty$ limit, for $b=1$, the constant (times $S_{\rm EE}$) that the R\'enyi entropy asymptotes to  is $\log \lambda$, where $\lambda$ is the largest eigenvalue of $\rho_{\rm v}$. Again we see that for $b>1$ (resp., $b<1)$ this constant gets reduced (enhanced).

Varying $b$ if $q$ is not fixed to 1,  or $q$ if $b$ is not fixed to 1, will reveal zeros in the denominator in the definition~(\ref{eq:general-entropy}) of $S_{q,b}$. This causes the whole function to  diverge at: 
\begin{equation}q=q_{\rm c}\equiv\frac{2}{(d-1)b^2-(d-3)}\ ,
\end{equation}
where for $b>1$, $q_{\rm c}<1$.  As an example, figure~\ref{fig:renyi-fixed-b} shows the ratio $S_{q,2}/S_{\rm EE}$ as a function of $q$ for $b=2$, for $d=3$. 
It is not clear if such a divergence has a direct physical interpretation. More mild discontinuities in  R\'enyi entropies   have been studied in this context in {\it e.g.,} refs.~\cite{Belin:2013dva,Belin:2014mva,Belin:2013uta}, but those 
 \begin{wrapfigure}{l}{0.45\textwidth}
\centering
\includegraphics[width=0.4\textwidth]{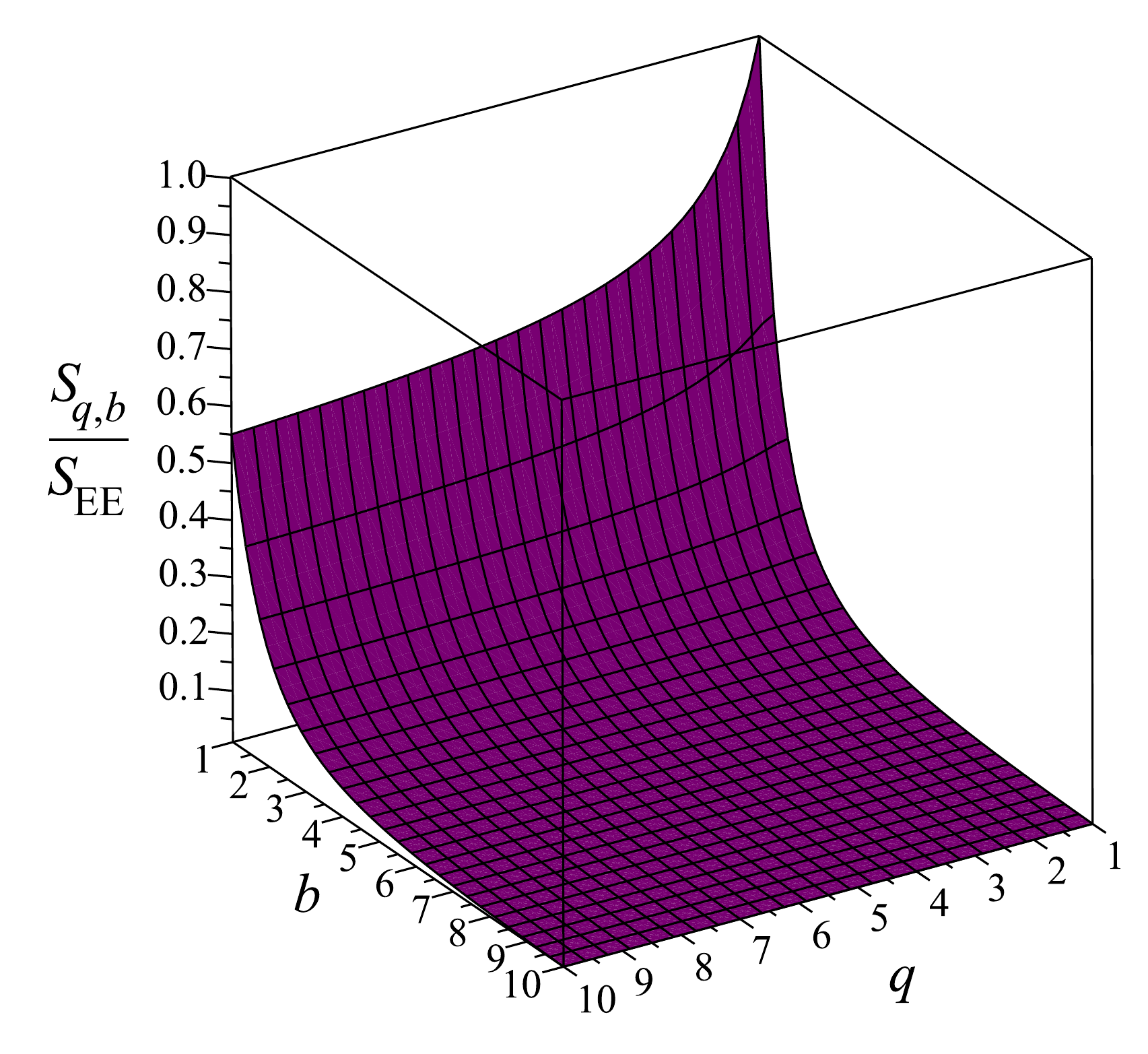}
\caption{\label{fig:S_q_b_3d} The ratio  $S_{q,b}/S_{\rm EE}$, {\it vs.} $q$ and~$b$, for $d=2$. The back walls show $S_{q,1}/S_{\rm EE}$ and $S_{1,b}/S_{\rm EE}$.}
\end{wrapfigure}
are connected to known examples of instabilities in the gravity (or gravity--plus--scalar) theory,  where masses cross a threshold, or a specific heat goes negative. In general, there does not seem to be such a physical   quantity associated with the denominator in eqn.~(\ref{eq:general-entropy}).

Note that, for $b>1$  the divergence is located at $q<1$, and so the region containing the entanglement $S_{1,1}$ and the ``higher'' entropies is smooth in~$q$. Figure~\ref{fig:S_q_b_3d} shows a three--dimensional plot of $S_{q,b}$ for $d=2$, for $1\leq \{q,b\}\leq10$. The shape  is also typical of the higher $d$ behaviour, for this range of parameters. It is also possible to entirely avoid the singularity  along one dimensional paths by making choices for $q$ and $b$ such that when either passes through unity, so does the other. $S_{q,1}$ and $S_{1,b}$ are two simple examples of such a choice, but there are others.  This leads to the third remark.

\begin{figure}[h]
\centering
\subfigure[]{\centering\includegraphics[scale=0.3]{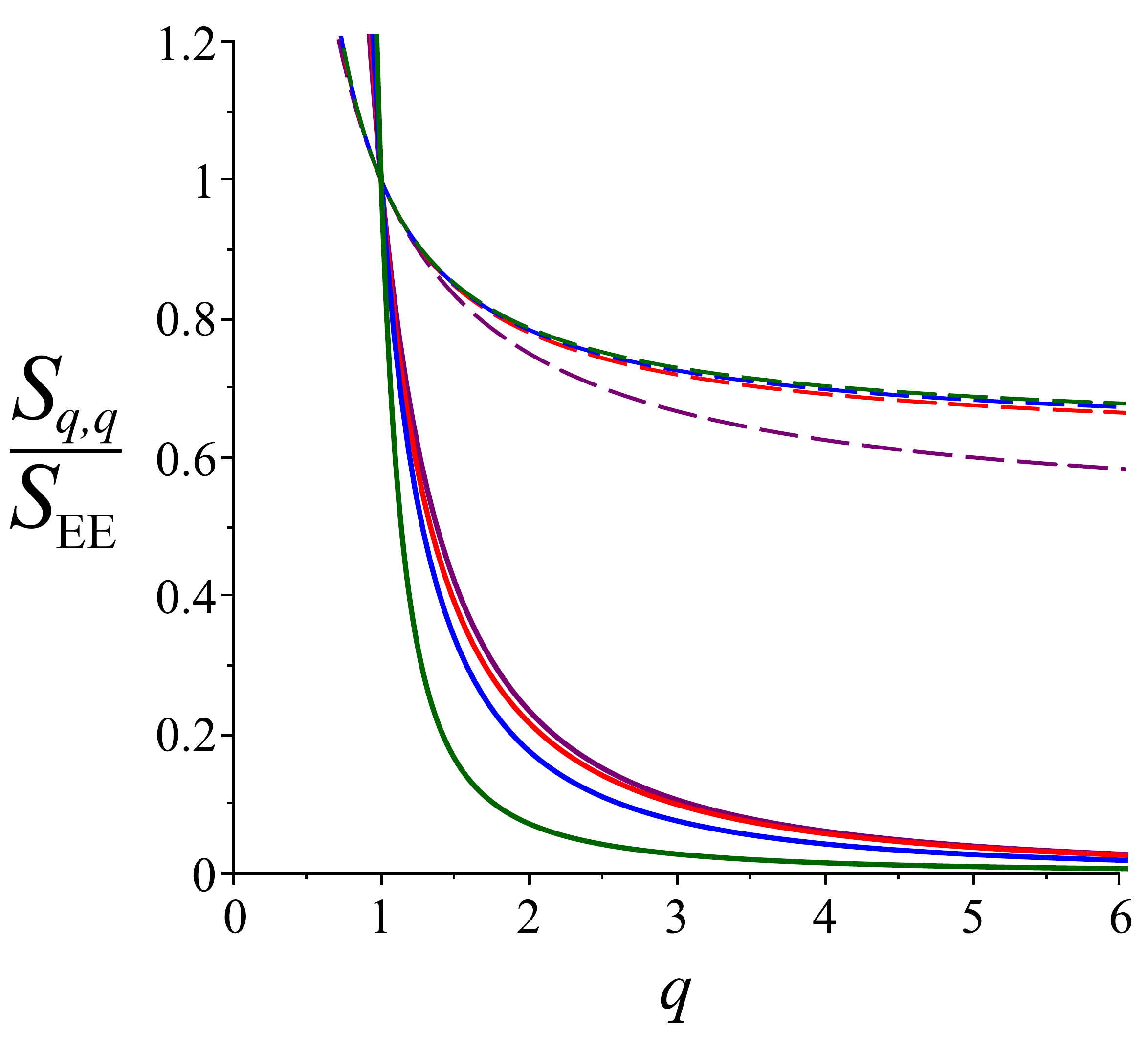}
\label{fig:renyi-qq}}
\hspace{30pt}
\subfigure[]{\centering\includegraphics[scale=0.3]{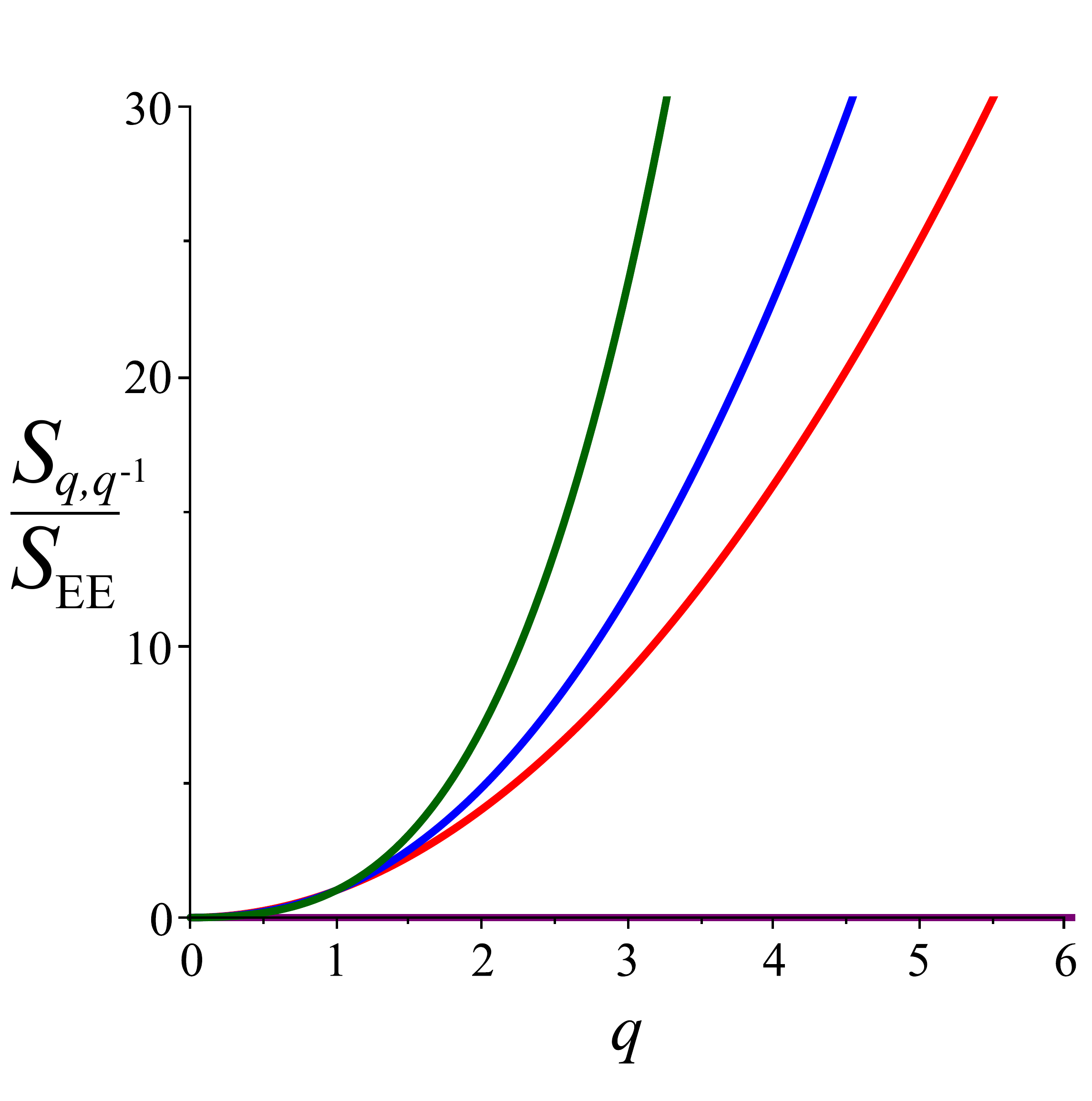}
\label{fig:renyi-qqinv}
}
\caption{The ratio $S_{q,b}/S_{\rm EE}$ as a function of $q$, when $b=b(q)$. (a) The $b=q$ case for $d=2,3,4$ and 10; (b) The $b=1/q$ case for $d=2$ (where it vanishes), $3,4$ and 5.}\label{fig:renyi-plots-b}
\end{figure}

{\bf (3)} The third remark is that it is natural (and also strongly motivated by the  previous remark) to  consider one dimensional paths in the $(b,q)$ plane  where the function $b(q)$ is  chosen such that~$b$ goes to unity with $q$.  The  dependence $b=q^p$ for various $p$ suggests itself, and we plot the case $p=1$ (so $S_{q,q}/S_{\rm EE}$) in figure~\ref{fig:renyi-qq}, which is sufficient to show the qualitative behaviour.  In this way we have an infinite family of new functions that fall off from unity at $q=1$, generalizing the R\'enyi behaviour (this is for $p>0$, the case of $p=-1$ appears in subsection~\ref{sec:special-cases}), but they all vanish as $2/(d-1)q^{2p}$ for large $q$ (see eq.~(\ref{eq:large-q})).

  There are many other interesting choices for $b(q)$ that can be made, and we will not attempt to classify them here. However,  two very special cases of a $q$--dependence for $b$ present themselves and have a natural interpretation. We discuss them next.

\subsection{Two Special Cases}
\label{sec:special-cases}
In figure~\ref{fig:pv-plane-b}, there are two special cases shown with dotted lines, labelled (i) and (ii). Let us study them in turn.
%
Case (i) is  when the displacement in pressure along the $T_0/q$ isotherm is exactly vertical, the exact counterpart to the pure R\'enyi case, which is horizontal. In that case we can see that the volume $V$ does not change, and so we still have $\rho_+=L_0$. Since $L=L_0/b$ we have $\bar{x}=\rho_+/L=b$. Some algebra shows that this results in $b^2-1=2(1-q)/qd$. Substituting this all into equation~(\ref{eq:general-entropy}) results in $S_{q,b}= S_{\rm EE}$. So we have found a deformation of R\'enyi that returns us to the Shannon/von Neumann case of entanglement entropy that is {\it different} from the $q,b\to1$ limit!  

There is a different way of understanding this result. Moves along vertical lines change neither $V$ nor $S$ since they are both simply powers of $\rho_+$. Therefore in this case $-\Delta G=S_0(\Delta T+V_0/S_0\Delta p)$, and so dividing by the normalization to get $S_{q,b}$ as defined in eq.~(\ref{eq:big-entropy}) gives,   after regularizing, the result. Finally, notice that this result covers vertical lines that extend to isotherms both below ($q>1$) and above ($q>1$) the $M=0$ curve of CFTs.

Case (ii) is when the displacement in pressure along the $T_0/q$ isotherm puts us back on the $M=0$ curve of conformal field theories. In this case, by definition, $\rho_+=L=L_0/b$ and hence $\bar{x}=1$. This results in the relation $b=1/q$, and
 \begin{equation}
S_{q,q^{-1}}= S_{\rm EE}\left(\frac{q(1-q^{d-2})}{(q-1)-q(d-1)(1-1/q^2)/2}\right)\ ,
\end{equation}
which vanishes for $d=2$. As before, this covers both $q>1$ cases (lower points on the $M=0$ curve---CFT deformations toward the IR) and $q<1$ cases (higher points---CFT deformations toward the UV). We shall revisit this observation in due course. The ratio $S_{q,q^{-1}}/S_{\rm EE}$ is plotted in figure~\ref{fig:renyi-qqinv} for $d=3,4$ and $d=5$. 

The unusual behaviour of the asymptotics of this family of curves fits with our earlier remarks. Recall that as $q\to0$, there is multiplication of  the $1/q^{d-1}$ behaviour by a $1/b^{2d-2}$ factor. In this case, $b=1/q$ and so this acts as a suppressing factor, sending that asymptotic to zero. In the case of the $q\to\infty$ regime, there's a $1/b^d$ factor as $b=1/q\to 0$.

\section{Field Theory Interpretation}
Next, it would be interesting to find out what this all means in the density matrix language that brought us here in the first place. We can attempt to do this by working backwards from the  thermodynamic presentation of subsection~\ref{sec:beyond}. 

\subsection{Generalities}
Starting with the first expression in eqn.~(\ref{eq:big-entropy}) we can use that 
\begin{equation}
G(T,p)=-T\log Z(T,p)=-T\log{\rm Tr}\left(e^{(-{\cal H}+pV)/T}\right)\ ,
\end{equation}
 to rewrite it as a difference of logarithms of two partition functions. A little further algebra gives:
\begin{equation}
S_{q,b}=\left(1+q\left(\frac{d-1}{2}\right)\left(\frac{b^2-1}{q-1}\right)\right)^{-1}\times \frac{1}{1-q}\log\left[\frac{Z(T_0/q,b^2p_0)}{Z(T_0,p_0)^q}\right]\ ,
\end{equation}
which suggests that
\begin{equation}
S_{q,b}=\frac{1}{[(1-q)-q(d-1)(b^2-1)/2]}\log\left[{\rm Tr}(\rho_{\rm v}^{(b)})^q\right] \ .
\end{equation}
In other words, generalizing the conformal map~(\ref{eq:conformal-to-thermal}) that connects the flat space CFT to the thermal ensemble on $\mathbb{R}\times\mathbb{H}^{d-2}$, we have:
\begin{equation}
\rho_{\rm v}^{(b)} = U^\dagger\left(\frac{e^{-({\cal H}+b^2p_0V_0)/T_0}}{Z(T_0,p_0)}\right) U
\end{equation}
where $\rho_{\rm v}^{(b)}$ is the density matrix that results from 
the addition of $\exp(-\Delta p V_0/T_0)$ inside the trace over states in the thermal description. The next question is what is $\rho_{\rm v}^{(b)}$ directly in the CFT. This is not fully clear for all $d$, but the idea that it is the $b$th power of another object suggests itself. 

The intuition behind this is as follows. Consider first the ordinary R\'enyi case of defining the $q$th power,~$\rho_{\rm v}^q$. To make $q$ copies of the density matrix $\rho_v$, represented as a thermal state at temperature~$T_0$, the system first needed to be fractionated into $q$ systems, each at temperature~$T_0/q$. They are then glued together (the next section reviews how this is done for $d=2$) to make  $\rho_{\rm v}^q$. Now we see that we have the same kind of machinery in place, but using the pressure variable $p_0$, equivalent to length scale $L_0$. By going to pressure $b^2p_0$,  it would seem that  
we are fractionating the system into $b$ copies, each with length scale $L_0/b$. Notice that, as explained in ref.\cite{Johnson:2018amj},  since this is a push into the IR, the degrees of freedom have been reduced by a factor of $b$ in each of the~$b$ copies. Presumably these copies are then glued together to make a system of scale $L_0$ again  and so our $\rho_{\rm v}^{(b)}$ must represent the $b$th power of one of these copies, which we'll call $\rho_{\rm w}$. So
\begin{equation}
\label{eq:new-powers}
\rho_{\rm v}^{(b)} = \rho_{\rm w}^{b}\ ,\quad {\rm with}\quad {\rm Tr}(\rho_{\rm w})= 1\ ,
\end{equation}
 where the latter condition seems appropriate for an individual density matrix. It leads to the prediction that for $b=1/q$, forming the $q$th power using the temperature sector gets cancelled out by the $(1/q)$th power coming from the pressure sector. Naively, this should yield a trivial result in this case. 
 We will be able to say  much more about  the $d=2$ case in  subsection~\ref{sec:d=2}, and see that this suggestion is fully realized there in terms of the properties of the CFT and the twist operators that perform the gluing. For higher $d$, we will also see some supporting evidence when we consider higher dimensional twist operators.

\subsection{The case of $d=2$: Twisted CFT}
\label{sec:d=2}
Let us examine the case of $d=2$ in a little more detail. There, we can write the full expression out succinctly, since ${\bar x}=1/qb$, and so:
\begin{equation}
\label{eq:d=2-renyi}
S_{q,b}=\frac{1}{[q-1+q(b^2-1)/2]}\cdot q\left(1-\frac{1}{q^2b^2}\right) \frac{c}{6}\log\left(\frac{\ell}{\epsilon}\right) \ ,
\end{equation}
where we have used the fact that $4\pi L_0/L_p=c/3$ where $c$ is the standard central charge in $d=2$. Also, the ``ball'' of radius $R$ is in this case a line segment of length $2R$, which we have denoted $\ell$. The first factor is just our standard normalization for our generalized quantity, designed to give the correct meaning to the $q,b\to 1$ case. However, the pre--factor of the logarithm has a direct interpretation in terms of a CFT computation. A brief review of the case of $b=1$ will be helpful in understanding it. (See {\it e.g.} refs.~\cite{Holzhey:1994we,Calabrese:2004eu,Calabrese:2005zw,Hung:2011nu,Klebanov:2011uf} for more details.) 

Start with $b=1$, and give  our $d=2$ CFT a spatial coordinate $x$ and a periodic time coordinate~$\tau$ with period $\beta=2\pi L_0=1/T_0$. Our interval is of length $\ell$ and  runs from $x=u$ to $x=v$. To implement the ``replica trick'' that computes ${\Tr} \rho_{\rm v}^q$, the theory is enlarged to the $q$--sheeted Riemann surface with coordinate $w=x+i\tau$, and our interval becomes a cut of length $\ell$ along the $x$--axis. We use a conformal map  complex coordinate $\zeta=(w-u)/(w-v)$ where $w=x+i\tau$. Here, the ends of the cut are at $0$ and $\infty$.  Two twist operators, $\sigma_q$ and $\sigma_{-q}$, (primary fields in the CFT) connect the~$q$ cuts by acting cyclically through them, ultimately connecting the first to the last. The quantity of interest, ${\Tr} \rho_{\rm v}^q$, turns out to be the correlation function $<\!\!\sigma_q(u)\,\sigma_{-q}(v)\!\!>$. 
In $d=2$ this correlator is, up to a constant, $|(u-v)/\epsilon)|^{-(h_q+h_{-q})}$. So the problem boils down to figuring out the twist fields' weights. Their weights are equal, and are:
\begin{equation}
\label{eq:b=1-conformal-weight}
h_q= h_{-q}=\frac{c}{12} q\left(1-\frac{1}{q^2}\right)\ ,
\end{equation}
the sum of which is precisely the pre--factor of the logarithm in the R\'enyi entropy. This value for the conformal weights is computed  by working on the single copy of the complex plane with  coordinate~$z$ arrived at by mapping our multi--sheeted Riemann surface thus: $z=\zeta^{1/q}$. The stress tensor components in each coordinate system are related by $T(w)=(dz/dw)^2 T(z)+(c/12)\{z,w\}$, where the last ``anomaly'' term contains the Schwarzian derivative:
\begin{equation}
\label{eq:schwarzian}
\{z,w\}\equiv\frac{(z^{\prime\prime\prime}-\frac32 (z^{\prime\prime})^2)}{(z^{\prime})^2}=\frac{1}{2}\left(1-\frac{1}{q^2}\right)\frac{(u-v)^2}{(w-u)^2(w-v)^2}\ .
\end{equation}
We can use translational invariance to fix $T(z)=0$, leaving this Schwarzian derivative as the key ingredient needed. The stress tensor $T^{(q)}(w)$ in the $q$--copy CFT is $q$ times $T(w)$, and the correlator  we need, $<\!\!\sigma_q(u)\,\sigma_{-q}(v)\!\!>$, is determined by inserting $T^{(q)}(w)$ into it and using a standard Ward identity for primary fields, with the result that their weights are indeed given by eqn.~(\ref{eq:b=1-conformal-weight}), following directly from eqn.~(\ref{eq:schwarzian}).

Now we turn to the $b\neq1$ case. We are still computing a $q$th power using the replica trick, but now it is of $\rho_{\rm v}^{(b)}$, the density operator of our system  for $b>0$. Recall that this means we've gone to higher pressure, meaning lower $L=L_0/b$. Nearly all of the details stay the same for computing the $q$--copy theory, and again the crucial result will all boil down to the correlator of the twist fields. It is their conformal weight that changes. The change comes about because the CFT we are working on has periodicity $L_0/b$, so the uniformizing  relation between coordinates must reflect that, and  is instead: $z=\zeta^{1/qb}$. This modifies the Schwarzian derivative by replacing the $1/q^2$ by $1/(qb)^2$, and after multiplying by $q$ again to get $T^{(q)}(w)$ for this theory, everything else goes through as before, showing that the conformal weights in the $b\neq1$ theory are:
\begin{equation}
\label{eq:bneq1-conformal-weight}
h^{(b)}_q= h^{(b)}_{-q}=\frac{c}{12} q\left(1-\frac{1}{q^2b^2}\right)\ ,
\end{equation}
the sum of which is precisely what appears in our generalized R\'enyi entropy eqn.~(\ref{eq:d=2-renyi}). As an independent conformation of this result, we'll show that this arises naturally in another approach in subsection~\ref{sec:d=2-holographic}.

It is amusing that for the $d=2$ theory the effect of $b$ is to simply modify the map from~$z$ to~$w$ through a  scaling.  It is for this reason that  when $b=1/q$ it precisely undoes the job of the uniformization map, giving $z=\zeta$ and causing the Schwarzian derivative to vanish,  sending the conformal weights of the twist operators to zero. The result is that $S_{q,q^{-1}}=0$ for this special case of $d=2$, precisely what we saw in case (ii) in subsection~\ref{sec:special-cases}. This fits extremely well with the interpretation of $b$ as being the number of copies of the density matrix $\rho_{\rm w}$ that the reduced vacuum is made of (see eqn.~(\ref{eq:new-powers}) and the comments just after).
This special feature that occurs in  $d=2$  a result of the pleasant conveniences afforded by conformal maps between Reimann surfaces does. not obviously persist for higher dimensions. On the other hand, if eqn.~(\ref{eq:new-powers}) is correct, the case $b=1/q$  ought to be special in any dimension, representing something about the system dramatically simplifying. We have seen that  $b=1/q$ is indeed a special solution in all dimensions, representing staying on the conformal  ($M=0$) line in the $(p,V)$ plane. We will also see in the next section that the weight of the higher $d$ analogs of the twist operators also vanish at $b=1/q$, supporting our expectation! Oddly, however, as we saw in section~\ref{sec:special-cases}, for $d>2$, $S_{q,q^{-1}}\neq0$.

Note that we can  understand case~(i) of section~\ref{sec:special-cases} here too. In this case of $d=2$, the relation between~$b$ and $q$ is $b=1/q^{\frac12}$, and the conformal weights sum to $\frac{c}{6}(q-1)$ while the normalizong denominator becomes $(q-1)/2$. So they cancel to the constant $c/3$ leaving us with $S_{\rm EE}=(c/3)\log(\ell/\epsilon)$. As with case (i) above, we see in $d=2$, through maps between Riemann surfaces, the simple way that the special value of $b$ changes the weight of the twist operators to adjust the theory.

\subsection{Holographic Computations for Twist Operators}
\label{sec:d=2-holographic}
In ref.\cite{Hung:2011nu}, a complementary (holographic) computation of the conformal weight~(\ref{eq:b=1-conformal-weight}) of the twist operator was presented in terms of gravitational quantities. In summary, the result can be written in terms of the energy density of the theory, ${\cal E}(T)$, as follows:
\begin{equation}
h_q=\frac{2\pi q L_0^d}{(d-1)}\left[{\cal E}(T_0)-{\cal E}(T_0/q)\right] = \frac{2\pi q }{\omega_{d-1}}\frac{L_0}{(d-1)}\left[M(T_0)-M(T_0/q)\right] \ ,
\end{equation}
where $M$ is in fact the black hole mass given in eqn.~(\ref{eq:the-mass}). Interestingly, once we interpret this in extended thermodynamics, we see that this is again another  difference of a state potential, this time the enthalpy, $H(S,p)=M$. Once again, as we saw with the Gibbs free energy, we see that the thermodynamics begs to have its natural structure fully used. Instead of staying at constant pressure, one can adjust it as well, obtaining, after a little algebra:
\begin{equation}
h_q^{(b)}= \frac{2\pi q }{\omega_{d-1}}\frac{L_0}{(d-1)}\left[M(T_0,p_0)-M(T_0/q,b^2p_0)\right] = q\left(\frac{L_0^{d-1}}{8G_{\rm N}}\right) {\bar x}^{d-2}(1-{\bar x}^2)\ ,
\end{equation}
where ${\bar x}$ is given by eqn.~(\ref{eq:general-entropy}). This would appear to be, for  $b\neq1$, our generalization  of the weight of the general $d$--dimensional ``twist" operators discussed in refs.\cite{Hung:2014npa}. In the case of $d=2$ something special happens. Our $S_{q,b}$ expression eqn.~(\ref{eq:general-entropy}) and our weight $h_q^{(b)}$ collapse to the same dependence on ${\bar x}$, in accordance with the fact that in $d=2$, $S_{q,b}$ is given in terms of the correlator, $<\!\!\sigma_q(u)\,\sigma_{-q}(v)\!\!>$, of the twist operators (see the previous subsection).  In this case, ${\bar x}=1/qb$ and  $L_0/8G_{\rm N}=c/3$, and we hence we recover the weight derived in eqn.~(\ref{eq:bneq1-conformal-weight}). The special case of $b=1/q$ corresponds to ${\bar x}=1$ and both the weight $h_q^{(b)}$ and $S_{q,b}$, being proportional to $(1-{\bar x}^2)$, vanish in accordance with expectations. 

Notice that for $d>2$ the solution ${\bar x}=1$ persists, still corresponding to the case $b=1/q$. This means that the higher dimensional twist operator weights vanish, reflecting the expected triviality  when one makes the $q$th power out of the $(1/q)$th power of $\rho_{\rm w}$. However, as already  remarked, although one would expect $S_{q,b}$ to vanish (because of the logarithm), it is not proportional to $(1-{\bar x}^2)$ for $d>2$. This deserves to be better understood. It is possible that there is a subtlety with defining $\rho_{\rm v}^{(b)}$, or perhaps with continuing $b$ away from integer values in  these cases.

\section{Concluding Remarks}
While there are many possible (and interesting) extensions, generalizations, and deformations of the R\'enyi entropy, the structures presented here are distinguished by being inspired by the elegant thermodynamic way of presenting the R\'enyi entropy\cite{2011arXiv1102.2098B}. In this sense they are quite natural to explore.  While the presentation of the extension in purely thermodynamic terms in section~\ref{sec:introduction} was interesting enough for an exploration (and is worth pursuing for its own sake in case there are applications to information theory or other fields), it is very satisfying to see that all of the elements needed are in place in extended gravitational thermodynamics, with an application to quantum information in conformal field theories in $d$ dimensions.

In the CFT the presence of $b$ seemed to amount to having the reduced density matrix of the vacuum be in a very special form: It is itself the $b$th power of a density matrix. It would be interesting to understand this better in the field theory,  for all $d$. Questions naturally arise as to when this is possible and/or useful in a given field theory, and the thermodynamic dual picture here is a natural setting in which to determine the answers. In a sense, we've found a refinement of the field theory structure that could amount to $S_{q,b}$, being  a  useful tool for studying more complex theories. 

As mentioned above and in section~\ref{sec:introduction}, it  would be interesting to explore whether our way of generalizing R\'enyi entropy has applications elsewhere. Moreover it would be interesting to understand further aspects of its properties, perhaps from an information theoretic perspective.   Key to doing so would be the identification of the analogues of $p$ and $V$, the pressure and volume variables, in the systems of interest.

\section*{Acknowledgements}
The work of CVJ  was funded by the US Department of Energy  under grant DE-SC 0011687.  CVJ would  like to thank  the Aspen Center for Physics for hospitality, and  Amelia for her support and patience.

\bibliographystyle{utphys}
\bibliography{renyi_entropy_deformation}

\end{document}